\RequirePackage{fix-cm} 
\documentclass[journal]{new-aiaa}
 \usepackage{mathtools}
 \usepackage{longtable,tabularx}
 \usepackage{varioref}
 \usepackage{wrapfig}
 \usepackage{threeparttable}
 \usepackage{dcolumn}
  \newcolumntype{d}{D{.}{.}{-1}}
 \usepackage{nomencl}
  \makeglossary
 \usepackage{subfigure}
 \usepackage{subfigmat}
 \usepackage{fancyvrb}
  \fvset{fontsize=\footnotesize,xleftmargin=2em}
 \usepackage{lettrine}
 \usepackage[dvips]{dropping}
 \usepackage[final]{pdfpages}
 \usepackage{setspace}
 \usepackage{amsmath}
 \usepackage{multicol}
 \title{A Morphing Continuum Simulation of Transonic Flow over an Axisymmetric Hill}
 \author{  Louis B. Wonnell,\footnote{AIAA Student Member, Graduate Student}
and Mohamad I. Cheikh, \footnote{AIAA student Member, Graduate Student}}

 \affil{Department of Mechanical and Nuclear Engineering\\
Kansas State University, Manhattan,\\ KS 66506, USA} 

 \author{James Chen \footnote{AIAA Senior Member, Assistant Professor, Corresponding Author, chenjm@buffalo.edu}}

 \affil{Department of Mechanical and Aerospace Engineering\\
University at Buffalo, The State University of New York,\\
Buffalo NY 14260, USA} 

 \newcommand{\eqnref}[1]{(\ref{#1})}

\begin{document}

\maketitle
\begin{abstract}
Finite volume simulations of turbulent boundary layer flow over an axisymmetric
hill are performed for $Re_{H} = 6500$ using Morphing
Continuum Theory (MCT), and compared with DNS data from Castagna et. al.
and experimental data obtained by
Simpson. The inlet profile was specified
by inputting values from the profile specified by Castagna et. al.
Root-mean-square velocity fluctuations are inputted using the new variable of
gyration from MCT. Additional terms introduced by MCT lead to a new formulation
of the Q-criterion, which allows for the visualization of three-dimensional
turbulent structures including a hairpin vortices. Streamline plots of the
separation bubble show a more confined bubble than Castagna et. al., but agree well on
the separation and
reconnection points. The surface pressure coefficient matches Simpson
much more closely than Castagna et. al. MCT data was obtained on a $6.72 \times 10^6$
cell mesh containing a smaller number of cells by an order of magnitude than
that of the DNS mesh of $5.4 \times
10^7$ by Castagna et. al. The effects of the particle on the
larger structures are
inferred from the variation in the new variable of gyration.
\end{abstract}

\section*{Nomenclature}
{\renewcommand\arraystretch{1.0}
\noindent\begin{longtable*}{@{}l @{\quad=\quad} 
l@{}}

$\alpha$ & material constant, $Pa*s*m^2$ \\
$\alpha_{1}$ & dimensionless parameter \\
$\alpha_{2}$ & dimensionless parameter \\
$\alpha_{3}$ & dimensionless parameter \\
$a_{kl}$ & deformation rate tensor, $1/s$ \\
$b_{kl}$ & deformation rate tensor, $1/(m*s)$ \\
$\beta$ & material constant $Pa*s*m^2$ \\
$c$ & speed of sound of the fluid, $m/s$ \\
$c_p$ & specific heat at constant pressure \\
$c_v$ & specific heat at constant volume \\
$C_p$ & pressure coefficient \\
$\delta$ & boundary layer thickness, $m$ \\ 
$\delta_{kl}$ & Kronecker delta tensor \\
$e$ & internal energy density, $(m/s)^2$ \\
$E$ & total energy density, $(m/s)^2$ \\
$\epsilon_{klm}$ & Levi-Civita tensor \\
$f_{k}$ & body force density, $m/s^2$\\
$\gamma$ & subscale diffusivity, $Pa*s*m^2$ \\
$\Gamma_{\phi}$ & diffusion coefficient, $m^2/s$ \\
$g$ & geometric interpolation factor \\ 
$h$ & energy source density, $m^2/s^3$ \\ 
$H$  & height of bump, $m$ \\
$i_{kl}$  & subscale inertia tensor \\
$j$  & microinertia, $m^2$ \\
$\kappa$ & subscale viscosity, $Pa*s$ \\
$l_{k}$ & body moment density, $(m/s)^2$ \\
$\lambda$ & second coefficient of viscosity, $Pa*s$ \\
$\mu$ & first coefficient of viscosity, $Pa*s$ \\
$m_{kl}$ & moment stress tensor, $Pa*m$ \\ 
$M_{\infty}$ & Mach number at inlet \\
$p$ & hydrostatic pressure, $Pa$ \\
$\phi$ & transport variable \\
$q_{k}$ & heat flux, $kg/s^3$ \\
$\rho$ & density, $kg/m^3$\\
$\sigma$ & thermal conductivity, $W/(m^2 K)$ \\
$S_{\phi}$ & source function for $\phi$ \\
$\textbf{S}_{f}$ & surface vector of the cell face \\ 
$t_{kl}$ & Cauchy stress tensor, $Pa$ \\
$\theta$ & temperature, Kelvin \\
$II_{a}$ & second invariant of $a_{kl}$ \\
$u^{+}$ & x-velocity in wall coordinates \\
$u_{rms}$ & root-mean-square velocity, $m/s$ \\
$u_{\tau}$ & friction velocity, $m/s$ \\
$V_{c}$ & cell volume, $m^3$ \\ 
$v_{k}$ & translational velocity vector, $m/s$ \\
$\omega_{k}$ & gyration vector, $1/s$ \\
$X_{k}$ & position vector for point P in reference configuration, m\\
$\xi_{k}$ & position vector from center at point P, m \\
$\chi_{kK}$ & directors attached to point P \\
$y^{+}$ & y-coordinate in wall units\\
$\omega_\text{rms}$ & troot-mean-square gyration, $1/s$\\
$N$ & number of sampling points\\
$\omega_i'$ & gyration fluctuation, $1/s$\\
$I$ & turbulence intensities\\
\end{longtable*}}

\section{Introduction}

\lettrine[lines=2]{T}{he} demand for comprehensive, robust, and efficient
numerical methods for modeling turbulence, while also yielding new physical
insights into well-studied cases, is a continuous challenge for each new model
\cite{wilcox1988reassessment, rodi1997comparison, park2004implementation,
pitsch2006large}\hspace{0.1mm}. The reliability of {\color{black} direct 
numerical simulation} (DNS) in
producing results that conform to experimental data is well documented, but
these results are obtained at the expense of considerable computational
resources \cite{van1956suggested, eringen:64ijes, joslin1993spatial,
furst2013comparison, chen2012numerical}\hspace{.1mm}. The primary difficulty is
the ability
to resolve all relevant scales of motion, and in particular the smallest
eddies. For these eddies, viscous diffusion of energy into heat is the primary
mode of energy transfer. Since this cascade process is how energy is
ultimately dissipated, no complete description of turbulent physics can omit
these scales of motion. In order to resolve these eddies in DNS, the mesh must
be continuously refined, driving up the computational costs.

Due to this hurdle, the Navier-Stokes equations have been
modified with averaging or filtering techniques, producing the familiar
large-eddy simulation (LES) and Reynolds-averaged Navier-Stokes (RANS) methods.
Though these techniques do obtain credible results at cheaper costs, they still
require closure models that change depending on the system being studied.
Typically, different versions of well-tested models are applied to canonical
problems to test their viability, cost, and potential for solving more complex
problems. Rodi et al's tests  of different versions of the $k-\epsilon$ model
for vortex-shedding flow illustrate the limits of various closure models and
the occasional necessity of combining different
models to approximate increasing levels of turbulence around a bluff
body\cite{rodi1997comparison}\hspace{.1mm}. Walters and Coljkat proposed a
three-equation
model that incorporated closure models for the RANS equations based on the
transport of different types of fluctuations to
different regimes of the flow \cite{walters2008three}\hspace{.1mm}. Depending
on the
nature of the results, or the geometry of the flow, different RANS models
may be utilized. These models do not come from any first principles and are
typically \textit{ad hoc}, based of particular experiments or
observations. They can be successful in adapting to new problems.
Still, the smallest relevant length scales in turbulent motion may be smoothed
over in the case of averaging, or filtered out as in the case of LES.

Developing a theory that captures these smallest length scales without
resorting to \textit{ad hoc} closure models is the goal of morphing continuum
theory (MCT) \cite{chen2012numerical,
heinloo2004formulation, kirwan1986boundary, mehrabian2008cosserat,
peddieson:72jes}\hspace{.1mm}. When the dynamics of a fluid's inner structure
are
considered, the resulting governing equations provide new terms that account
for the motion or deformation of individual eddies. These terms arise from
the independent local rotation of finite-sized particles in the fluid that give
the fluid its inner structure. Recent successes in numerical simulations of
two-dimensional boundary layer flows and compressible
turbulent flows have produced visualizations of shock-turbulent interactions,
turbulent kinetic energy transfer, transitional profiles, and other experimental
fluids phenomena \cite{chen2015multiscale, wonnell2016morphing,
wonnell2017morphing}\hspace{.1mm}. The key to these simulations is the
ability to generate data consistent with experiment and DNS results,
without the associated high computational costs \cite{chen2012numerical,
chen2015multiscale, wonnell2016morphing, wonnell2017morphing}\hspace{.1mm}.
Additionally,
the new variable of local rotation, the new material properties associated
with the inner structure, and the new governing equations allow for alternative
strategies of visualizing large structures within the flow. These tools can
help with the visualization of
three-dimensional turbulent boundary layer flow. To prove MCT's applicability
to physical cases of turbulence, the theory must be able to
capture the structures involved in three-dimensional turbulence. {\color{black} MCT should not
be confused with RANS or LES. RANS and LES both work
by taking the existing Navier-Stokes equations
and adding approximations or modifications to make the simulation of turbulent flows
more cost-effective.
In those theories, the fluid is modeled as a classical continuum with infinitesimal points.
In MCT, the fluid is modeled as a continuum with finite-sized inner structures possessing rotation. The rotation is similar to the rotation of a solid body, which can be solved 
independently with the conservation of angular momentum. In RANS or LES, an arbitrary length scale, e.g. integral length, or turbulence model must be used to close the system of governing equations. In MCT, these equations flow directly from the nature of
continuum mechanics and the second law of thermodynamics.} This paper
will derive new, objective, mathematical tools from MCT for visualizing
structures within the three-dimensional flow.

The aim of this paper is to demonstrate the capability of MCT to produce
compressible, three-dimensional, turbulent flow data that demonstrate at least
qualitative consistency with experimental and numerical results. In
addition, the effectiveness of the objective Q-criterion will be tested
in its ability to visualize structure in three-dimensional turbulence. Section
\ref{MCTTurbulence} presents the mathematical principles, variables,
parameters, and derivations leading to the governing equations. From these
governing equations,
an objective invariant, similar to the Q-criterion in
the Navier-Stokes equations, is derived.  Section \ref{numericalimplementation}
will deal with the conversion of the analytical equations to a form suitable
for numerical application. The numerical values for the material properties
will be
introduced, with their values determined by key dimensionless parameters.
Section \ref{results} will produce the
results from the three-dimensional compressible axisymmetric hill simulation
and compare them with experimental and numerical data. Contour plots of the
new, frame-indifferent Q-criterion will unveil large structures such as hairpin
vortices within the flow.
Finally, section \ref{Conclusion} will present concluding remarks on the
reliability of the new MCT data, the value added by the new
objective Q-criterion, and the future of MCT in modeling three-dimensional
turbulence.

\section{Morphing Continuum Theory and Turbulence} \label{MCTTurbulence}

The classical continuum is characterized as a collection of infinitesimal points while
the morphing continuum is defined as a continuous collection of finite size
{\color{black} particles}.
The Navier-Stokes equations
solve for the fluid profile at infinitesimal points throughout the domain, with
the solutions describing the motion and interaction of groups of particles.
With Eringen's introduction of a new independent variable of sub-scale motion,
the influence of the motion of {\color{black}small particles } on the macroscopic 
behavior of
the flow can be described mathematically\cite{eringen:66jmm}\hspace{.1mm}.
\begin{wrapfigure}{L}{.4\linewidth}
 \includegraphics[width=1\linewidth]{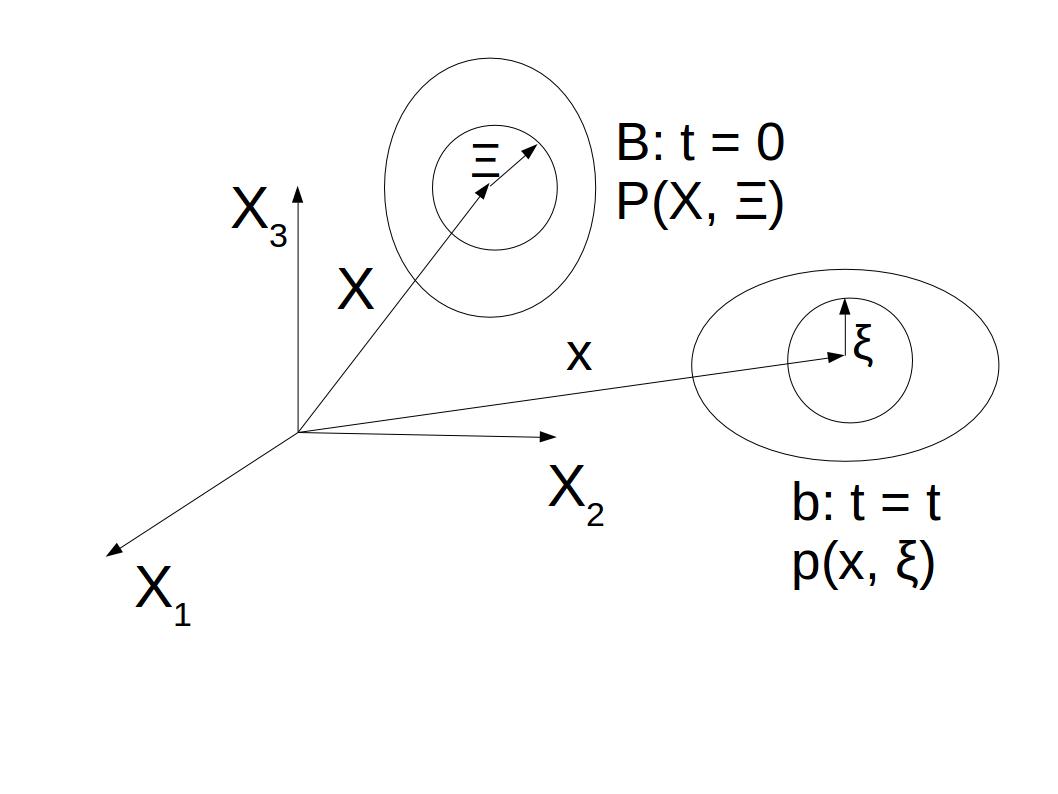}
 \caption{Evolution of general structure with associated sub-scale motion
vector
\textbf{$\xi$} and macromotion vector
\textbf{x}}
 \label{f:mctdirectors}
\end{wrapfigure}
Local differences in the properties of the fluid are now
explicitly stated through the addition of new properties related to the
particles. While some turbulence models introduce terms such as eddy
viscosity, Eringen's theory derives new material properties related to the
sub-scale viscosity and diffusivity \cite{eringen:66jmm}\hspace{.1mm}. The
importance of these parameters in generating turbulence in the flow will be
explained later in this section.

Individual particles in a morphing continuum can behave similarly to 
eddies in a turbulent flow. Figure \ref{f:mctdirectors} depicts the 
simultaneous sub-scale motion of an
eddy by a local vector $\xi_{k}$ and a director $\chi_{kK}$ to account for any
deformation of this eddy. These vectors constantly influence and react to the
macroscopic variables. Therefore, any forces that drive the overall behavior of
the fluid must also alter the sub-scale motion vectors. Additionally, the
motion of these vectors is shown to be a one-to-one mapping, meaning the
vectors satisfy relations of the form:
\begin{align}\label{e:direc1}
 x_{k} &= x_{k} (X_{K},t)
\quad
 &X_{K}& = X_{K} (x_{k},t)\\
 \xi_{k} &= \chi_{kK}(X_{K},t)\Xi_{K}
\quad
 &\Xi_{K}& =  \bar{\chi}_{Kk}\xi_{k}
\end{align}
Here, $x_{k}$ and $X_{K}$ refer to the macroscale vector
components of the position of a particle, with initial coordinate indices
denoted by $K$ and final directions denoted by $k$. Equation \ref{e:direc1}
adds nine more degrees of freedom of
sub-scale motion to the classical fluids description, making a concise picture
of the overall behavior of the fluid a challenge. In the case of Morphing
Continuum Theory, only three extra degrees of freedom will be introduced due to
the assumption of rigid directors (c.f. equation \ref{e:orthonormal}).

The relationship between these individual particles in the fluid and
eddies in a turbulent flow is worth addressing. Richardson notes that the
range of eddy sizes within turbulence is
arbitrary\cite{richardson:22cup}\hspace{.1mm}.
One can only approximate an upper limit to eddy size based on a problem's
largest relevant length scale. Furthermore, Richardson notes, turbulent
structures blend together, making the tracking of eddy sizes a challenge. As
Richardson notes, "big whirls have lesser whirls and so on to viscosity- in the
molecular sense" \cite{richardson:22cup}\hspace{.1mm} In other words, if a
large eddy
appears to form in a turbulent flow, further investigation will reveal that this
eddy is in fact composed of smaller eddies, which themselves may be composed of
even smaller eddies. This observation requires any analytical theory to start
from a carefully selected small length scale and reconstruct larger eddies from
that scale.

{\color{black} Morphing continuum theory } achieves this goal by assuming a 
quality of the fluid
known as sub-scale isotropy. This assumption allows for the elimination of the
extra degrees of freedom of sub-scale motion mentioned earlier. Any elongations
or compressions of the eddies in a turbulent flow are presumed to be isotropic
at the length scale of the smallest eddies. As a mathematical consequence, the
directors used to describe sub-scale motion become orthonormal
\cite{eringen:66jmm}\hspace{0.1mm}:
\begin{equation}\label{e:orthonormal}
\chi_{kK}\chi_{lK} = \delta_{kl}
\quad
\chi_{kK}\chi_{kL} = \delta_{KL}
\end{equation}
In short, any stretching motion of a particle is irrelevant, and the only
motion that becomes relevant at this smallest scale is its local rotation.
Deformations of any larger structures in the flow still occur, as in
classical fluids. For individual particles, equation \ref{e:orthonormal} leaves
only sub-scale rotation and translation. Each particle is
characterized as a rigid sphere with self-spinning rotation. Since individual
particles can represent individual eddies, larger eddies can always be composed
of eddies with their own independent rotation. Richardson's picture becomes a
mathematical description of the fluid.

With the introduction of sub-scale deformations, Eringen defines a property
corresponding to a particle's resistance to that
deformation\cite{eringen:66jmm}\hspace{.1mm}:
\begin{equation} \label{e:microinert}
{\it j_{kl} \equiv} {\it i_{mm}\delta_{kl}-i_{kl}}
\end{equation}
The $i_{kl}$ term serves a role similar to the moment of inertia of the
particle, and has its own definition\cite{chen2012micropolar}\hspace{.1mm}:
\begin{equation} \label{microinert1}
{\it i_{kl} \equiv \frac{\int_{\delta v'}\rho'(\chi, \xi, t)\xi_{k}\xi_{l}dv'}{\int_{\delta v'}\rho'(\chi, \xi, t)dv'} \equiv <\xi_{k}\xi_{l}> }
\end{equation}
Taking the average of the trace of the subscale inertia tensor
from equation~\eqnref{e:microinert} yields:
\begin{equation} \label{constmicroinert}
{\it j \equiv \frac{j_{mm}}{3}}
\end{equation}
As discussed earlier, MCT uses spherical particles as the
fundamental components of the fluid. These spheres still contain their
own inertia, with the characteristic length scale being the diameter $d$ of
the sphere. Chen showed that a fluid of spherical particles leads to the
relation $j = \frac{2}{5}d^2$\cite{chen:12caf}\hspace{.1mm}. Therefore, $j$ is
a parameter
available for controlling the size of eddies. If the smallest eddy remains
smaller than the smallest element within a mesh, then this parameter allows for
the modelling of that subgrid scale. For this
paper the fluid will consist of a set of particles with uniform size,
thus making the parameter $j$ constant.

With the addition of the rotation of the particles, the concept of
strain in the fluid must be altered from that of classical fluids. Taking
material derivatives of the deformation tensors in equation \ref{e:direc1},
Eringen derives the morphing continuum strain-rate tensors, $a_{kl}$ and
$b_{kl}$, as\cite{eringen:66jmm}\hspace{.1mm}:
\begin{equation}
\label{Astrain}
{\it a_{kl}  = v_{l,k} + \epsilon_{lkm}\omega_{m}}\\
\end{equation}
\begin{equation}
\label{Bstrain}
{\it b_{kl} = \omega_{k,l}}\\
\end{equation}
These new tensors show how the gyration, $\omega_{m}$, or local rotation of the
particles adds to the familiar strain-rate, which is produced in classical
fluids only by transverse velocity gradients. The $b_{kl}$ tensor indicates
that an entirely new strain is experienced from simple gradients in the
gyration. The physical meaning of this strain will become evident when stress
tensors are expressed in terms of the deformation tensors.
With this new variable for gyration, MCT provides the means of calculating
the rotational kinetic energy of individual structures. Given a fluid density
$\rho$, the rotational energy of a particle then becomes $\frac{1}{2}\rho j
\omega_{k}\omega_{k}$. Since these particles mimic the behavior of
eddies, this expression can be used to approximate turbulent kinetic energy.
In the simulations of turbulence before and after a shock, this
term will carry the information about eddy energy transfer
\cite{wonnell2016morphing, cheikh2017morphing}\hspace{0.1mm}.

The governing equations for MCT are derived from balance laws for mass, linear
momentum, angular momentum, and energy. In tensor form, the balance laws are
written as:
\begin{equation}
\frac{\partial \rho}{\partial t} + (\rho v_{i})_{,i} = 0
\label{Continuity}
\end{equation}
\begin{equation}
{\it t_{lk,l} }+ \rho{\it f_{k} } =  \rho {\it \dot{\it v}}_{\it k} \\
\label{LinMomentum}
\end{equation}
\begin{equation}
{\it m_{lk,l} + \epsilon_{kij}t_{ij}} + \rho{\it l_{k}} = \rho{\it j} {\it \dot{\omega}}_{\it k} \\
\label{AngMomentum}
\end{equation}
\begin{equation}
\rho \dot{e} - t_{kl}a_{kl} - m_{kl}b_{lk} + q_{k,k} - \rho h = 0
\label{Energy}
\end{equation}
Dotted terms refer to their material derivatives, i.e. $\dot{v}_{k}= 
(\frac{\partial}{\partial t} + v_{l}\frac{\partial }{\partial x_{l}}) 
v_{k}$. Here, $l_{k}$ is the body moment density, $t_{kl}$ is the Cauchy stress 
tensor,
$f_{k}$ is the body force density, $m_{kl}$ is the moment stress tensor, $e$
is the internal energy density, and $q_{k}$ is the heat flux. These laws are
derived directly from rational continuum thermomechanics (RCT) and apply to any
morphing continuum, the term used for any space composed of spherical inner
structures \cite{eringen:66jmm, eringen:64ijes, eringen:99spring,
eringen:01spring}\hspace{.1mm}.

To obtain the governing equations for MCT, constitutive
equations are used to relate the Cauchy and moment stresses and the heat flux to
the velocity and gyration in the flow. These linear constitutive equations are
derived to be \cite{eringen:66jmm, chen:12caf,
chen2011constitutive}\hspace{.1mm}:
\begin{equation}
\label{CauchyStress}
{\it t_{kl} = -p\delta_{kl} + \lambda tr(a_{mn})\delta_{kl} + (\mu + 
\kappa)a_{kl} + \mu a_{lk}}
\end{equation}
\begin{equation}
\label{MomentStress}
{\it m_{kl} = \frac{\alpha_{T}}{\theta} \epsilon_{klm} \theta_{,m} + \alpha tr(b_{mn})\delta_{kl} + \beta b_{kl} + \gamma b_{lk}}
\end{equation}
\begin{equation}
q_{k} = -\frac{\sigma}{\theta}\theta_{,k} +
\alpha_{T}\epsilon_{klm}\omega_{m,l}\\
\label{heatflux}
\end{equation}
Here $p$ is the pressure, $\mu$ is the first coefficient of viscosity,
$\lambda$ is
the second coefficient of viscosity, $\epsilon_{klm}$ is the permutation tensor,
$\kappa$ is the sub-scale viscosity coefficient, $\gamma$ is the
sub-scale diffusivity, $\alpha$ and $\beta$ are material constants, $\it
\theta_{,m}$ is the temperature gradient, $\sigma$ is the thermal conductivity,
and $a_{kl}$ and $b_{kl}$ are the deformation-rate tensors shown earlier. When
the expressions for deformation-rate tensors in equations \ref{Astrain} and
\ref{Bstrain} and stress tensors are plugged into the balance laws with body
forces neglected, the governing equations have the form:
\begin{equation}
{\it \dot{\rho} + \rho v_{k,k} } = 0\\
\label{VecContinuity}
\end{equation}
\begin{equation}
{\it -P_{,k} + (\lambda + \mu)v_{l,lk} + } \\
       {\it (\mu + \kappa)v_{k,ll} + \kappa(\epsilon_{klm}\omega_{m,l}) =
\rho\dot{v}_{k} }
\label{VecLinMomentum}\\
\end{equation}
\begin{equation}
{\it (\alpha + \beta)\omega_{l,lk} + \gamma \omega_{k,ll}
+ \kappa(\epsilon_{klm}v_{m,l} - 2\omega_{k})  = \rho j\dot{\omega}_{k} }\\
\label{VecAngMomentum}
\end{equation}
\begin{equation}
(t_{kl}v_{l})_{,k} + (m_{kl}\omega_{l})_{,k} - q_{k,k} + \rho h = \rho\dot{E}\\
\label{VecEnergy}
\end{equation}
where $E = e + \frac{1}{2}(v_{l}v_{l} + j\omega_{l}\omega_{l})$ is the total
energy density of the fluid. The dotted variables refer to the material derivative of
those variables.

To close this system of equations, the pressure and density must be related
to the specific energy $e$. For this case, the fluid is presumed to be an
ideal gas, leading to the relations:
\begin{equation}
e = c_{v}\theta = c_{v}\frac{p}{\rho(c_{p} - c_{v})}
\label{idealgas}
\end{equation}
\begin{equation}
\rho E = \frac{p}{(\frac{c_{p}}{c_{v}} - 1)} + \frac{1}{2}\rho(v_{l}v_{l} + j\omega_{l}\omega_{l})\\
\label{idealGasEnergy}
\end{equation}
Here, $c_{p}$ and $c_{v}$ refer to the specific heats of the fluid at constant
pressure and volume respectively. The ratio $\frac{c_{p}}{c_{v}}$ was set to the
value for air, $1.4$.

These equations set the foundation for the numerical solver and the process of
visualizing eddies in the flow. Before the details of the numerical
analysis can be explained, a key mathematical tool from the governing equations
must be derived.

\subsection{The Q-criterion of MCT}

The subject of the identification of a vortex has been discussed for decades,
with various mathematical and physical arguments given for how to isolate true
vortices in the flow \cite{jeong1995identification, hunt1988eddies,
chen2015comparison, haller2005objective}\hspace{.1mm}. Qualitative definitions
involving
swirling motion or quantitative arguments based on the presence of high
vorticity were among the first criteria proposed to distinguish vortices in the
flow. As defined by Hunt, the Q-criterion focused on restricting the value of the
velocity-gradient tensor and requiring the existence of a local minimum in
the pressure \cite{hunt1988eddies}\hspace{.1mm}.

This practice of determining which facets of the flow counted as true
vortices continued with the study of invariants. The local pressure minimum
criterion eventually evolved into a requirement that the well-known
$\lambda_{2}$ invariant remain negative in the flow. In these cases, vortices
become apparent for steady Navier-Stokes flows, with the observer sitting in a
Gallilean coordinate system \cite{haller2005objective}\hspace{.1mm}. Haller
noted, however,
that simple rotations of the coordinate system caused both the Q and
$\lambda_{2}$ criteria to break down.
Haller's
discussion expanded into a broader investigation of what kind of criteria
succeeded in identifying vortices, independent of the physical situation and
choice of coordinate system. Specifically, the invariants used to determine the
presence of a vortex needed to preserve ``objectivity,'' or indifference to a
precise change in the reference coordinate system
\cite{haller2005objective, Chen2018PRE}\hspace{.1mm}.

One invariant that satisfies this criterion emerges from the deformation-rate
tensor, $a_{ij}$,
defined in equation \ref{Astrain} \cite{eringen:64ijes}. This
strain
encompasses the traditional strain-rate produced from the velocity gradient and
adds the local rotation supplied by the gyration of individual particles.
Including both of these motions is essential to calculating objective
quantities in MCT. Eringen demonstrates that objectivity is maintained in a
morphing continuum by taking into account macroscopic and sub-scale
deformations simultaneously, both of which are featured in equation
\ref{Astrain} \cite{eringen:64ijes}\hspace{.1mm}. Only by accounting for these
motions
can $a_{ij}$ remain indifferent to coordinate transformations. This
characteristic of the deformation in MCT is essential for maintaining a rigorous
standard for vortex identification. This paper will seek to reformulate the
Q-criterion as a first step towards establishing an objective criterion for
identifying a vortex.

From linear algebra, one invariant calculates the level of
asymmetry in a two-dimensional matrix, e.g. the strain-rate tensor. Setting
this invariant to be $II_{a}$, the expression for this value has the form
\cite{jeong1995identification}\hspace{.1mm}:
\begin{equation}
\label{invariant}
 II_{a} =  \frac{1}{2}(a_{ii}a_{jj} - a_{ij}a_{ji})
\end{equation}
where $a_{ij}$ is the deformation-rate tensor. If the fluid obeys the
continuum assumption, then this expression expresses a balance
between the magnitude of the vorticity and the shear strain rate
\cite{jeong1995identification}\hspace{0.1mm}. Since the deformation-rate tensor
in equation
\ref{Astrain} contains the shear rate and the gyration, this new invariant
involves a balance of more aspects of the flow.

If the expression for $a_{ij}$ from equation \ref{Astrain} is substituted into
equation \ref{invariant}, the invariant becomes:
\begin{equation}
 \label{MCTinvariant}
 II_{a} = \frac{1}{2}[(v_{i,i})(v_{j,j}) - (v_{j,i} +
\epsilon_{jim}\omega_{m})(v_{i,j} +
\epsilon_{ijl}\omega_{l})]
\end{equation}
Multiplying and canceling the appropriate terms yields:
\begin{equation}
 \label{MCTinvariantsimplified}
 II_{a} = \frac{1}{2}[v_{i,i}v_{j,j} - v_{j,i}v_{i,j} -
2v_{j,i}\epsilon_{ijm}\omega_{m}
+ 2\omega_{m}\omega_{m}]
\end{equation}
where the identities $\epsilon_{jim} = - \epsilon_{ijm}$ and
$\epsilon_{jim}\epsilon_{ijl} = -2\delta_{ml}$ have been employed. Letting all
the indices go to their respective coordinates in the
three-dimensional Euclidean coordinate system, it is clear that the diagonal
terms associated with the compressibility of the fluid disappear when the
indices from the shear terms equal one another. Therefore, the invariant used
for vortex visualization in MCT under Euclidean coordinates has the form:
 \label{MCTfinalinvariant}
\begin{align}\nonumber
 II_{a} &= [v_{x,x}v_{y,y} + v_{x,x}v_{z,z} + v_{y,y}v_{z,z} -
(v_{x,y}v_{y,x} + v_{x,z}v_{z,x} + v_{y,z}v_{z,y}) \\ 
&- (v_{y,x} -
v_{x,y})\omega_{z} - (v_{x,z} - v_{z,x})\omega_{y} - (v_{z,y} -
v_{y,z})\omega_{x} + \omega_{x}^2 + \omega_{y}^2 + \omega_{z}^2]
\end{align}

The final version of the invariant contains the expected shear strain-rate
components from both the large and small scales of motion. The familiar
velocity gradient terms, contained in the Q-criterion for the Navier-Stokes
equations, are now balanced with the magnitude of the gyration for individual
particles. If the gyration terms disappear, the Navier-Stokes
Q-criterion is recovered \cite{jeong1995identification,
hunt1988eddies}\hspace{.1mm}.
This fact is a consequence of the governing equations of MCT
reducing to the Navier-Stokes equations when the gyration is set to zero. Also,
if the gyration equals zero and the macroscopic stresses cancel one another
out, then the macroscopic flow simply has no preferred rotational direction.
Finally, if all stresses become zero then, naturally, no vortical motion could
possibly exist and the invariant dissapears once more. As observed
by Jeong et al for the Navier-Stokes' Q-criterion, a non-zero value is expected to produce visible eddies
but may not be sufficient to visualize a true
vortex \cite{jeong1995identification}\hspace{.1mm}. Future investigations into
this criterion and how it relates to other formal definitions of vortices are
needed to make a conclusive case that MCT identifies true vortices. Still, the
new expression from MCT allows for a new objective tool of
visualizing eddies and turbulent structures of various sizes in the
three-dimensional flow.

\section{Numerical Implementation}\label{numericalimplementation}

\subsection{Finite Volume Method}

For this work, the finite volume method is used to discretize
the MCT balance laws. The spatial domain is divided into
connected control volumes, or cells, with the physical variables solved at the
center of each cell.

The transport equation for any conserved property can be written in the
following form:
\begin{equation}
\frac{\partial \phi}{\partial t} + \nabla \cdot (\textbf{v} \phi) - \nabla
\cdot (\Gamma_\phi \nabla \phi ) = S_\phi
\label{te}
\end{equation}
Here, $\phi$ refers to the transport variable, $\Gamma_\phi$ is the diffusion
coefficient, and $S_\phi$ is the source function for $\phi$. If $\phi = \rho$,
equation \ref{te} becomes the continuity equation \ref{VecContinuity}.
Setting $\phi = \rho v_i$ yields the linear momentum equation
\ref{VecLinMomentum}, while $\phi = j \rho \omega_i$ yields the new governing
equation for the angular momentum \ref{VecAngMomentum}. Finally, if $\phi = \rho
E$ the new energy equation \ref{VecEnergy} is obtained.

The finite volume method requires that the governing equations in their
integral form be satisfied over the control volume. Applying spatial
integration to Eq. \ref{te} gives:
\begin{equation}
\int_V \frac{\partial \phi}{\partial t} dV + \int_V  \nabla \cdot (\textbf{v}
\phi)
dV - \int_V \nabla \cdot (\Gamma_\phi \nabla \phi) dV = \int_V S_\phi dV
\end{equation}
The diffusion term in equation \ref{te} can be approximated
as:
\begin{equation}
\int_V \nabla \cdot (\Gamma_\phi \nabla \phi) dV = \int_S (\Gamma_\phi \nabla
\phi)
\cdot d \textbf{S} \approx \sum_f (\Gamma_{\phi } \nabla \phi)_f \cdot
\textbf{S}_f
 \end{equation}
where $\textbf{S}_f$ represents the surface vector of the face, and $\sum_f$
denotes the summation over the faces of the control volume. The term
$(\Gamma_\phi \nabla \phi)_f$ can be obtained from the weighted average of the
gradients at the centroids of the face multiplied by the diffusivity at
the centroid:
\begin{equation}
(\Gamma_\phi \nabla \phi)_f = g (\Gamma_\phi \nabla \phi)_O + (1-g)(\Gamma_\phi
\nabla \phi)_N
\end{equation}
where the subscripts $O$ and $N$ represent the nodes at the center of the
owner cells and neighbor cells respectively. Here, $g$ is the geometric
interpolation factor related to the position of the element face $f$ with
respect to the cell center $O$. The gradient term $\nabla \phi$ located at the
cell center is computed using the Green-Gauss theorem
\cite{darwish2015coupled}\hspace{.1mm}:
\begin{equation}
(\nabla \phi)_c = \frac{1}{V_c} \sum_f \phi_f \textbf{S}_f
\end{equation}

The nonlinear convection term in equation \ref{te} requires a special treatment.
The scheme adopted
for the convection term should be able to capture any shock waves and
discontinuities, and at the same time avoid oscillations. The convection
term can be approximated as:
\begin{equation}
\int_V \nabla \cdot (\textbf{v} \phi) dV = \int_S (\textbf{v} \phi) \cdot
d\textbf{S}
\approx \sum_f \textbf{v}_f \phi_f \cdot \textbf{S}_f
\end{equation}
Notable methods found in the literature that are able to effectively produce
accurate non-oscillatory solutions are:  piecewise parabolic method (PPM)
\cite{colella1984piecewise}\hspace{.1mm}; essentially non-oscillatory (ENO)
\cite{shu1988efficient,  harten1987uniformly}\hspace{.1mm}; weighted ENO (WENO)
\cite{liu1994weighted}\hspace{.1mm}; and the Runge-Kutta discontinuous Galerkin
(RKDG) method \cite{cockburn1998runge}\hspace{.1mm}. All of these methods
involve Riemann solvers,
characteristic decomposition and Jacobian evaluation, making them troublesome to
implement.

In the present solver a simple forward-Euler scheme is implemented for the
unsteady term:
\begin{equation}
\int_V \frac{\partial \phi}{\partial t} dV = \frac{\phi_c^n -
\phi_c^o}{\bigtriangleup t} V_c
\end{equation}
where $V_c$ represents the cell volume, the subscript $c$ gives the cell
center, and superscripts $n$ and $o$ refer to the new and old time values
respectively. In these schemes, different temporal solvers can be substituted.
One can input a higher-order Runge-Kutta time integration scheme to
achieve a higher level of accuracy. Kurganov et al demonstrated that stability
is achieved with the implementation of a modified Euler
method \cite{kurganov2001semidiscrete}\hspace{.1mm}. For this scheme, the
diffusion terms
in the momenta equations are solved at the new time step, meaning that the
implicit character that allows for numerical stability of the KNP algorithm is
preserved. The scheme implemented in this study is a second-order semi-discrete,
non-staggered scheme, introduced by Kurganov, Noelle and Petrova (KNP)
as a generalization of the Lax-Friedrichs scheme
\cite{kurganov2001semidiscrete}\hspace{.1mm}. The interpolation
procedure from the cell center to the face center implemented in this scheme is
split into two directions corresponding to the outward or inward direction of
the face normal:
\begin{equation}
\sum_f \textbf{v}_f \phi_f \textbf{S}_f  = \sum_f \left[ \alpha
\textbf{S}_{f+} \textbf{v}_{f+} \phi_{f+} + (1-\alpha)\textbf{S}_{f-}
\textbf{v}_{f-} \phi_{f-}+ \boldsymbol\omega_f(\phi_{f-} + \phi_{f+})  \right]
\label{KNP}
\end{equation}
where $\textbf{S}_{f+}$ is the same as $\textbf{S}_f$ and $\textbf{S}_{f-} = -
\textbf{S}_f$. The subscript $f+$ is denoted for the directions coinciding with
$\textbf{S}_{f+}$, and $f-$ for the opposite direction, $\alpha$ the weighted
coefficient, and $\omega_f$ is the diffusive volumetric flux. The two terms $
\textbf{S}_{f+} \textbf{v}_{f+} \phi_{f+} $ and $ \textbf{S}_{f-}
\textbf{v}_{f-} \phi_{f-} $ in equation \ref{KNP} represent the flux evaluated
at the $\textbf{S}_{f+}$ and $\textbf{S}_{f-}$ directions respectively. The
last term, $\boldsymbol{\omega}_f(\phi_{f-} + \phi_{f+})$, is an additional
diffusive term
based on the maximum speed of propagation of any discontinuity that may exist at
the face.
The weighting coefficient $\alpha$ is based on the local speed of
propagation shown below:
\begin{align}
\psi_{f+} = max \left( c_{f+} |\textbf{S}_f| + \phi_{f+}, c_{f-} |\textbf{S}_f|+
\phi_{f-}, 0 \right) \\
\psi_{f-} = max \left( c_{f+} |\textbf{S}_f| - \phi_{f+}, c_{f-} |\textbf{S}_f|-
\phi_{f-}, 0 \right)
\end{align}
Here $c$ is the speed of sound of the fluid. The weighting factor is:
\begin{equation}
\alpha = \frac{\psi_{f+}}{\psi_{f+} + \psi_{f-}}
\end{equation}
and the diffusive volumetric flux is:
\begin{equation}
\boldsymbol \omega_f =  \alpha (1 - \alpha) (\psi_{f+} +
\psi_{f-})
\end{equation}
\subsection{Initial and Boundary Conditions}

\begin{figure}[ht!]
\centering
 \includegraphics[width=75mm]{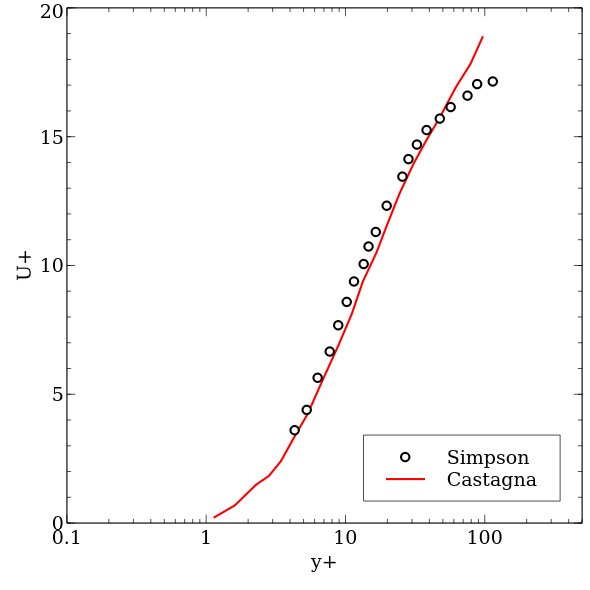}
 \caption{Comparison of DNS inlet profile obtained by Castagna et. al. 
\cite{castagna2014direct} from a precursour simulation with the experimental
profile used by Simpson \cite{simpson2002study}\hspace{.1mm}.}
 \label{inflowprofile}
\end{figure}

\begin{figure}[ht!]
\centering
 \includegraphics[width=75mm]{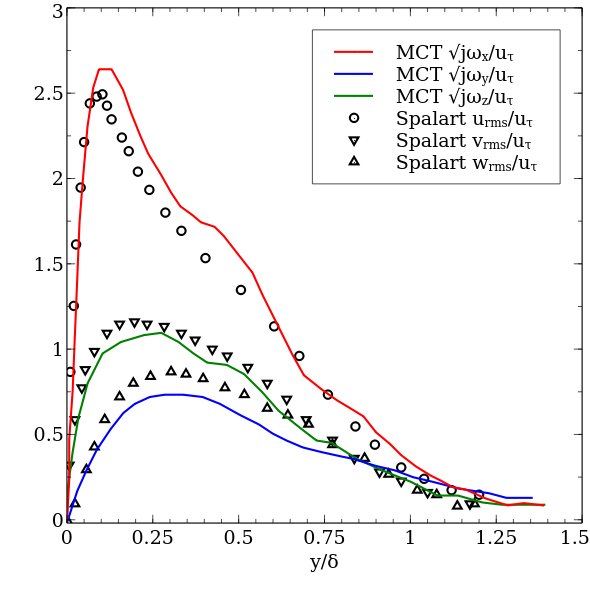}
 \caption{Comparison of MCT/DNS velocity perturbations the DNS
profile obtained by Spalart \cite{spalart1988direct}\hspace{.1mm}.}
 \label{inflowgyration}
\end{figure}

An inlet boundary layer profile with a thickness of $\delta = 0.039 m$ was
specified with an otherwise uniform flow of $M_{\infty} = 0.6$ at the inlet of
a $20H \ \times \ 3.205H \times \ 10H$ domain with $H = 0.078 m$ as per
Castagna et. al. \cite{castagna2014direct}\hspace{.1mm}. The inflow velocity profile
was specified by using the mean velocity of the precursor simulation done by Castagna et. al. \cite{castagna2014direct} and comparing the profile with the inflow
profile measured experimentally by Simpson
\cite{simpson2002study}\hspace{.1mm}.
Figure \ref{inflowprofile} shows that the inflow profiles match very well
near the wall and overpredict the velocity in the log layer where $y^{+} > 60$.
The profiles share a boundary layer thickness of $\delta = \frac{H}{2} =
0.039m$. This discrepancy was kept to see if MCT could still capture
experimental flow phenomena in the bulk flow. Turbulent fluctuations were
specified by equating the \textcolor{black}{root-mean-square (rms)} of the new variable of gyration with the root-mean-square (rms)
velocities specified by Castagna et. al. and compared with Spalart
\cite{castagna2014direct, spalart1988direct}\hspace{.1mm}. Figure
\ref{inflowgyration} compares the normalized root-mean-square data from MCT and Castagna et. al. with the experimental data obtained
by Spalart \cite{spalart1988direct}\hspace{.1mm}. The discrepancies are the
same for both
MCT and Castagna et. al. , since the data by Castagna et. al. was directly transferred to the
gyration variable, via the expression $\sqrt{j}\omega_{i,rms} = u_{i, rms}$.
Therefore, the kinetic energy generated by fluctuations in the velocity 
profile by Castagna et. al. matches the rotational kinetic energy produced by the gyration. 
Discrepancies between
the DNS and experimental data were most notable in the log-layer. These
discrepancies were preserved in the gyration, again to see if MCT could still
produce better agreement with experiment given the same initial flow data.
\begin{figure}[ht!]
\begin{subfigmatrix}{2}
 \subfigure[Internal Mesh]{\includegraphics[width=65mm]{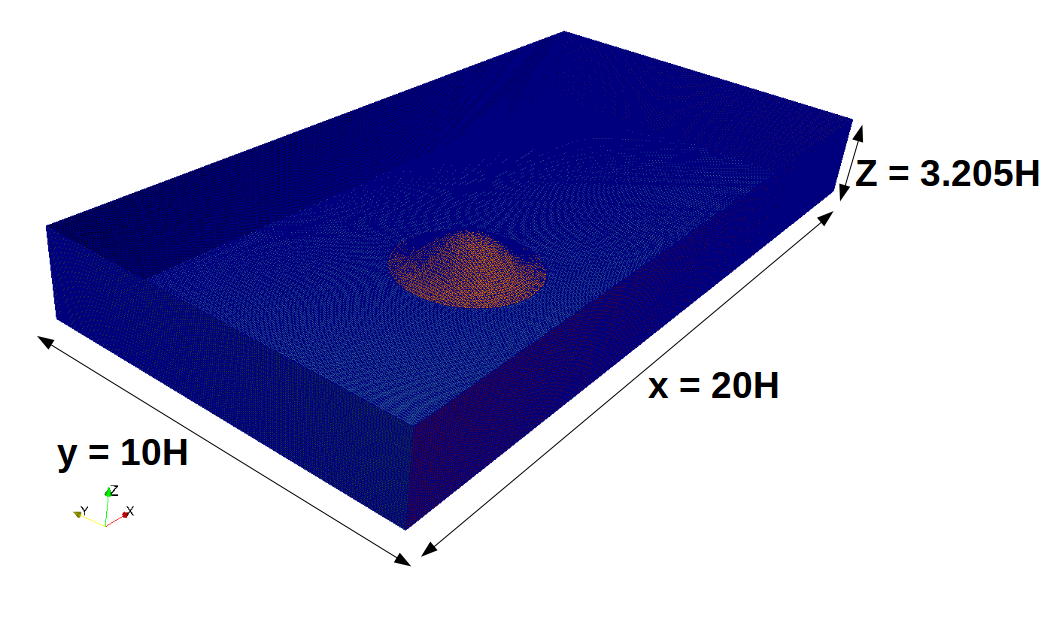}}
 \subfigure[Hill Mesh]{\includegraphics[width=75mm]{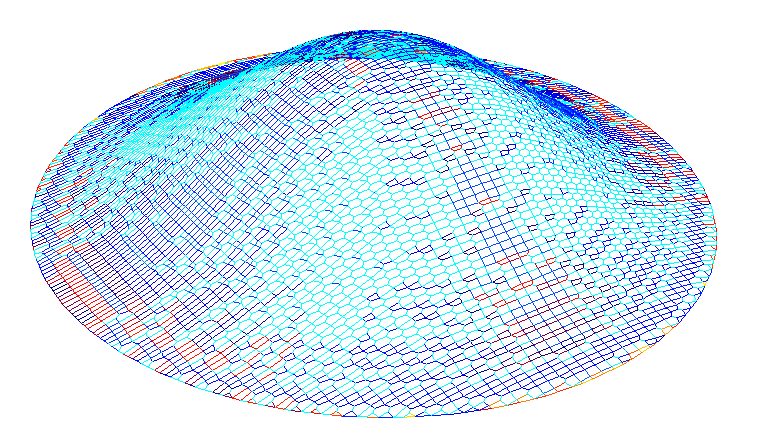}}
 \end{subfigmatrix}
 \caption{Wireframe of the meshes for the rectangular domain and hill.
Axisymmetric hill was set at 8.4H away from the interior.}
 \label{f:InternalMesh}
\end{figure}
Inflow density, total viscosity $\mu + \kappa$, and freestream
velocity $U_{\infty}$, were all set to ensure the Reynolds number, $Re_{H}$,
based on the height $H$ matched the value of 6500 by Castagna et. al. {\color{black} The 
time-step for the MCT simulations was $5 \times 10^{-8}$ s, shorter than the value of $3.04 \times 10^{-5}$ s by Castagna et. al..} {\color{black} The smaller mesh cells near the 
wall required these small time step values, but the resulting data benefited 
from increased temporal accuracy. The wall-normal mesh distribution was set to have 
a simple grading whereby the last cell away from the wall would be 
200 times the length of the smaller cell near the wall.} Non-reflective
boundary conditions were set at the top and outlet of the domain to prevent
unphysical pressure effects from affecting the dynamics at the hill and to
follow the setup by Castagna et. al.
\cite{castagna2014direct}\hspace{.1mm}. Periodic boundary conditions were set in
the
spanwise direction, also in line with the conditions by Castagna et. al.
\cite{castagna2014direct}\hspace{.1mm}. Zero gradient, and no-slip
boundaries for the velocity, were set at the floor and hill. The mesh near the
hill was tailored
to the shape of the hill, determined by the mathematical functions set in
Castagna et. al. \cite{castagna2014direct}\hspace{.1mm}. Table \ref{t:MeshStats} lists
the key
statistics associated with this custom mesh. A key note is the number of
cells for this mesh, totaling around 6.7M elements. {\color{black} With the 
unstructured mesh, however, the
viscous sublayer, defined by the limit $z^{+} < 10$, contained 30 cells as
opposed to the 10 cells needed in the mesh by Castagna et. al.
\cite{castagna2014direct}\hspace{.1mm}.}
Still, the
argument that MCT can provide results comparable to DNS data without the
associated computational costs is supported by the
dramatic decrease in mesh cell number.

\begin{table}[t!]
\centering
 \begin{tabular}{||c  c||}
 \hline
 Parameter & Value \\  [.5ex]
 \hline
 Maximum Aspect Ratio & 8.62886 \\
 Time Step & $5 \times 10^{-8}$ s \\
 Maximum Skewness & 5.124 \\
 $\Delta z^{+}_{min}$ & \color{black} 0.148 \\
 Number of cells: $z^{+} < 10$ & \color{black} 30 \\
 Total Number of Cells & $6.72 \times 10^6$ \\ \hline
 \end{tabular}
\caption{Parameters for mesh quality and time resolution used in MCT
simulations}
\label{t:MeshStats}
\end{table}

\subsection{Material Properties}

Three non-dimensonal parameters to gauge the onset of turbulence are introduced
by Peddieson \cite{peddieson:72jes} and are later explained by Wonnell and Chen
\cite{wonnell2016morphing}\hspace{.1mm}. These parameters can be extracted from
the
governing equations through dimensionless analysis
\cite{peddieson:72jes}\hspace{.1mm}. For
incompressible flow over a flat plate, these parameters produced turbulent
velocity profiles within a boundary layer that matched
experimental data produced by the European Research Community on Flow,
Turbulence and Combustion (ERCOFTAC) \cite{chen2015multiscale}\hspace{.1mm}.
The parameters are defined as follows:
\begin{equation} \label{nodimension}
{\it \alpha_{1} = \frac{\kappa}{\mu},\hspace{2 mm} \alpha_{2} = \frac{\kappa}{\rho\sqrt{j}U},\hspace{2 mm}\alpha_{3} = \frac{\gamma}{\mu j} }
\end{equation}
In the flat plate study, $\alpha_{1}$ proved to be the pivotal parameter in
matching an experimental turbulent profile
\cite{chen2015multiscale}\hspace{.1mm}.
This parameter serves as a ratio between the particles' contribution to the
Cauchy stress, $\kappa\epsilon_{klm}\omega_{m,l}$  and the classical viscous
diffusion term, $\mu v_{ll}$, in the linear momentum equation
\ref{VecLinMomentum}. Local variation in the gyration of the particles
leads to a tension in the fluid that disrupts the otherwise smooth laminar
flow. The classical viscous diffusion attempts to smooth disruptions
created by differences in gyration, and so the balance of these forces is
critical for determining whether a flow has reached a turbulent state. This
result indicates that the tension created by differences in rotational motion
of particles needs to exceed viscous diffusion by a considerable amount in
order to maintain turbulence within an incompressible boundary layer.

Table \ref{t:material} gives the values of the three dimensionless parameters
for $\alpha_{n}$ that successfully generated turbulence in the incompressible
case.  The problem of capturing sub-grid length scales becomes more important
when the turbulence becomes more compressible, as the smallest eddies could be
impacted by density fluctuations. The balance of compressibility with viscous
fluctuations occurs at all scales of motion. This balance is reflected in
the new total viscosity, $\mu + \kappa$, of the fluid found in the Reynolds
number, $Re_{H}$. The contribution of individual structures to the total
viscous resistance of the fluid is captured through $\kappa$. This simulation
incorporates no sub-grid models and will allow for the effects of compressible
turbulence to be taken into account.
\begin{table}[ht!]
\centering
 \begin{tabular}{||c  c||}
 \hline
 Parameter & Value \\  [.5ex]
 \hline
 $\alpha_{1}$ & 99 \\
 $\alpha_{2}$ & 0.0014 \\
 $\alpha_{3}$ & 0.235 \\
 $M_{\infty} = \frac{U_{\infty}}{c} $      & 0.6\\
 $Re_{H} = \frac{\rho_{\infty} U_{\infty}H}{\mu + \kappa}$ & 6500\\
\hline
 \end{tabular}
\caption{\small Dimensionless parameters $\alpha_{n}$, Mach Number $M$, and
the Reynolds number matching DNS  \cite{castagna2014direct, simpson2002study,
chen2015multiscale}\hspace{.1mm}.  Speed of sound determined for air at
$T_{\infty} = 293 K$}
\label{t:material}
\end{table}
\section{Results} \label{results}
Results were obtained after the freestream flow made 1.2 trips through the
domain to follow the example of Castagna et. al., or around $t = 0.009s$
\cite{castagna2014direct}\hspace{.1mm}. 
\begin{figure}[ht!]
\centering
\includegraphics[width=.7\linewidth]
{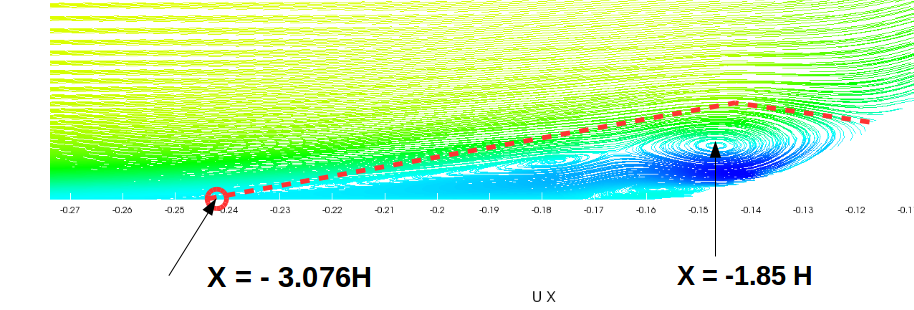}
\caption{Streamline compared with separation bubble boundaries obtained by
Castagna et. al. \cite{castagna2014direct}\hspace{.1mm}. MCT data demonstrate a 
larger windward side separation bubble, but no significant separation on the 
leeward side. MCT bubble delineated by red dotted line.}
 \label{f:SeparationBubble}
\end{figure}

\begin{figure}[ht!]
\centering
 \includegraphics[width=.35\linewidth]{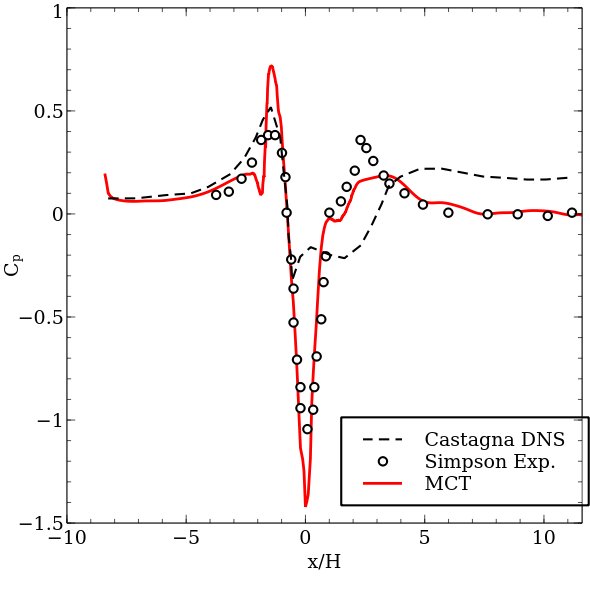}
 \caption{Comparison of $C_p = \frac{p_{static}
- p_{atm}}{.5*\rho*U_{\infty}^2}$
between experimental data from Simpson, simulation data from Castagna et. al., and
numerical data along the centerline
$z = 0$ \cite{simpson2002study, castagna2014direct}\hspace{.1mm}.}
 \label{f:SimpsonComparison}
\end{figure}
\begin{figure}[ht!]
\begin{subfigmatrix}{6}
 \centering
\subfigure[Streamwise Turbulence Intensity: x = 4.14H]
{\includegraphics[width=.3\linewidth]{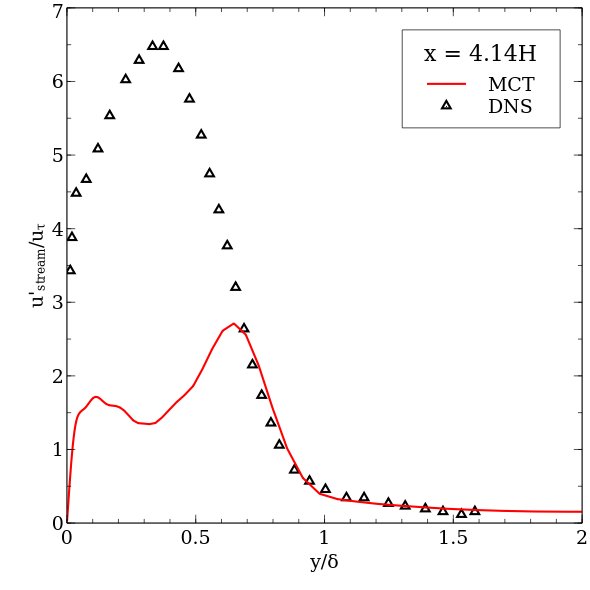}}
\subfigure[Spanwise Turbulence Intensity: x = 4.14H]
{\includegraphics[width=.3\linewidth]{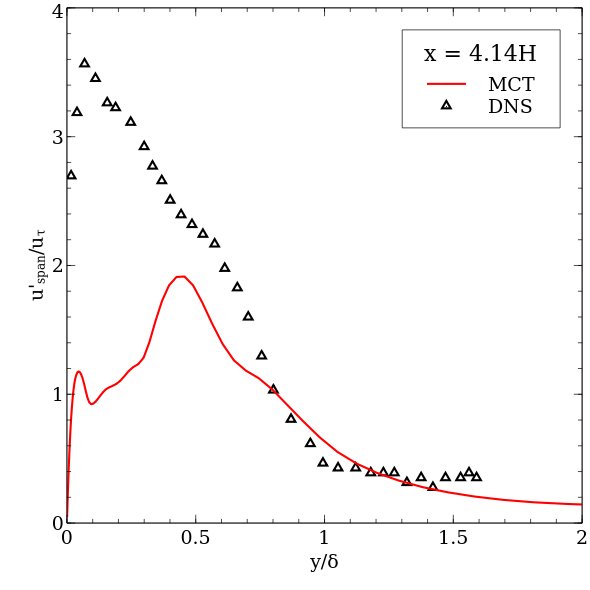}}
\subfigure[Wall-normal Turbulence Intensity: x = 4.14H]
{\includegraphics[width=.3\linewidth]{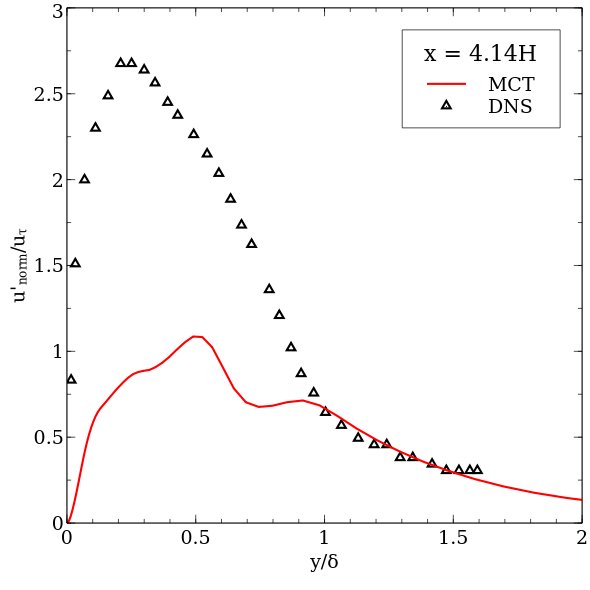}}
\subfigure[Streamwise Turbulence Intensity: x = 11.6H]
{\includegraphics[width=.3\linewidth]{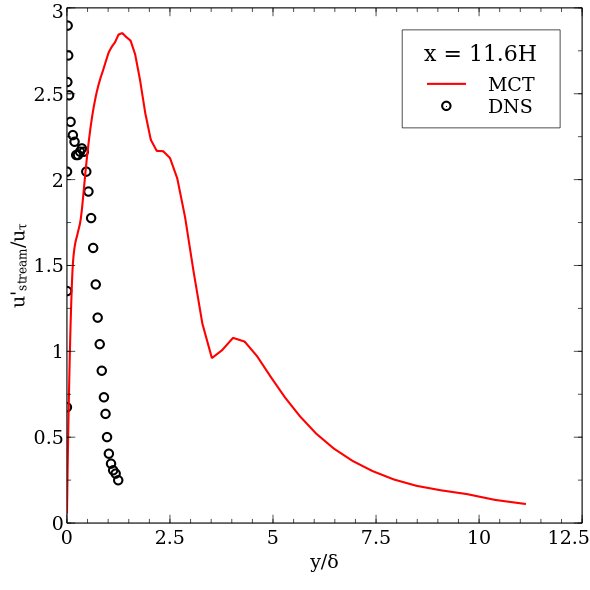}}
\subfigure[Spanwise Turbulence Intensity: x = 11.6H]
{\includegraphics[width=.3\linewidth]{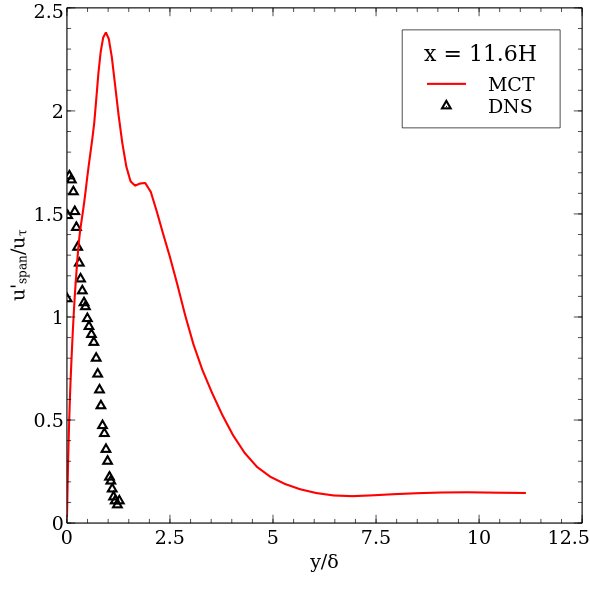}}
\subfigure[Wall-normal Turbulence Intensity: x = 11.6H]
{\includegraphics[width=.3\linewidth]{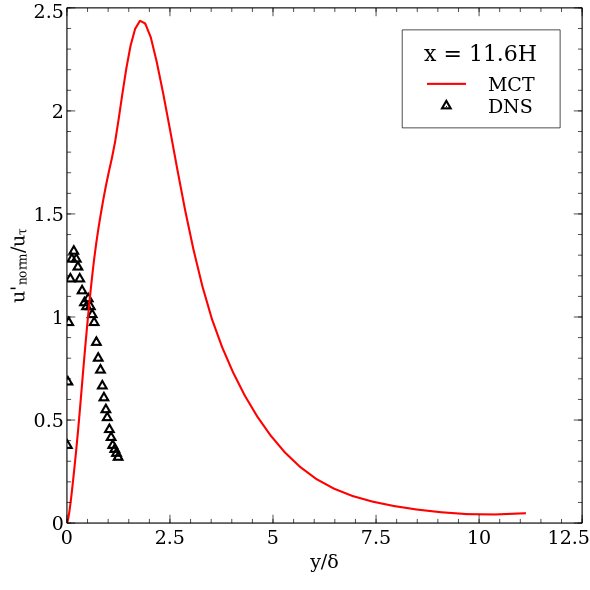}}
\end{subfigmatrix}
 \caption{Comparison of turbulence intensities, averaged in the spanwise
direction, at $x = 4.14H$ and $x = 11.6H$ with DNS data obtained by Castagna et. al. 
\cite{castagna2014direct}\hspace{.1mm}.}
\label{f:Gyr}
\end{figure}
\begin{figure}[ht!]
\begin{subfigmatrix}{3}
 \centering
\subfigure[Hairpin
Vortices: t = 0.016001 s]
{\includegraphics[width=75mm]{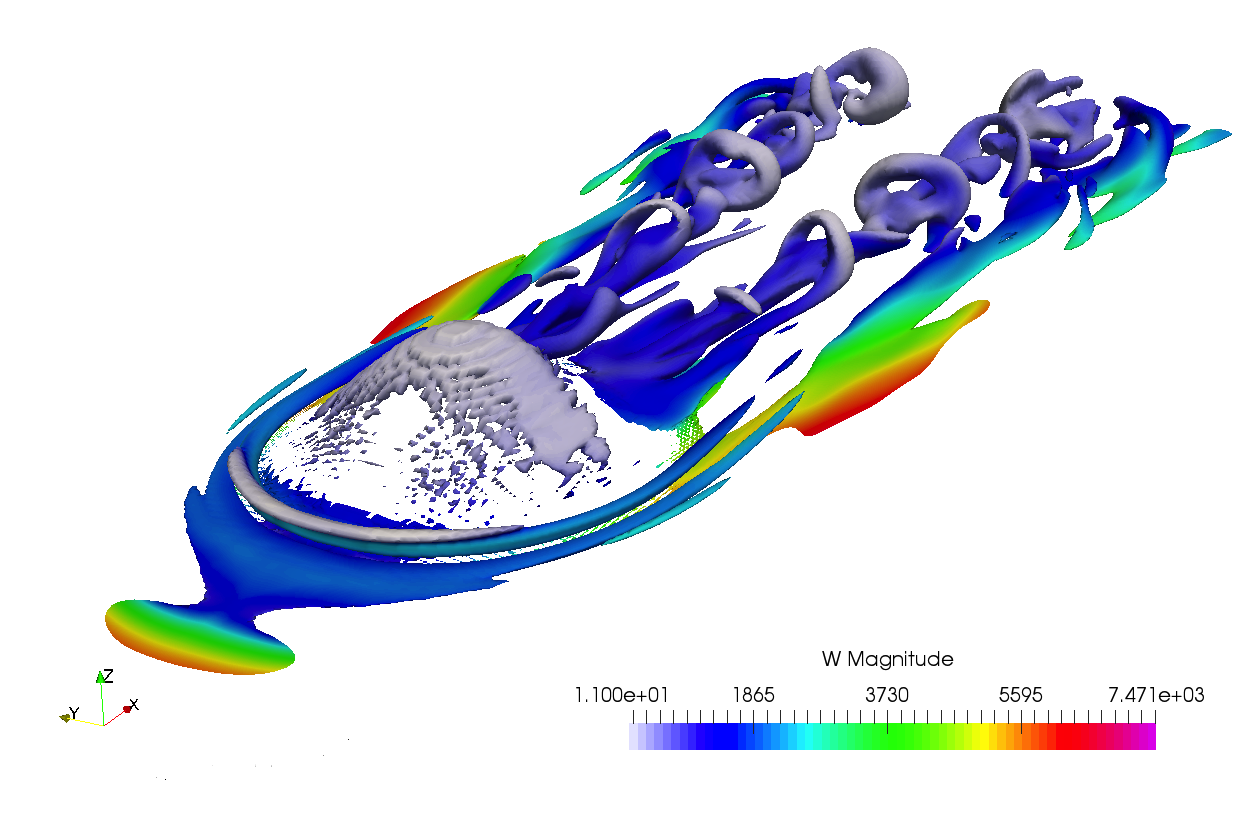}}
\subfigure[Hairpin Vortices: t = 0.016105 s]
{\includegraphics[width=75mm]{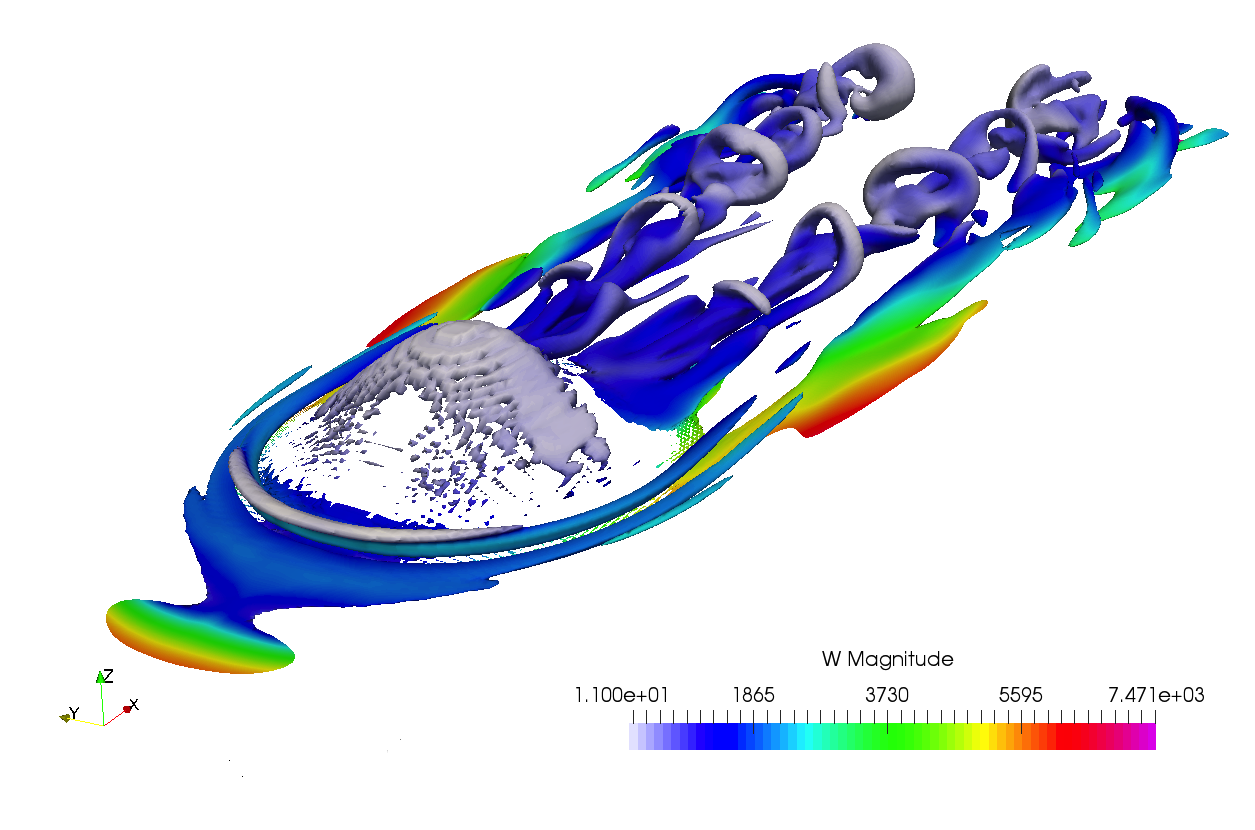}}
\subfigure[Hairpin Vortices: t = 0.016210 s]
{\includegraphics[width=75mm]{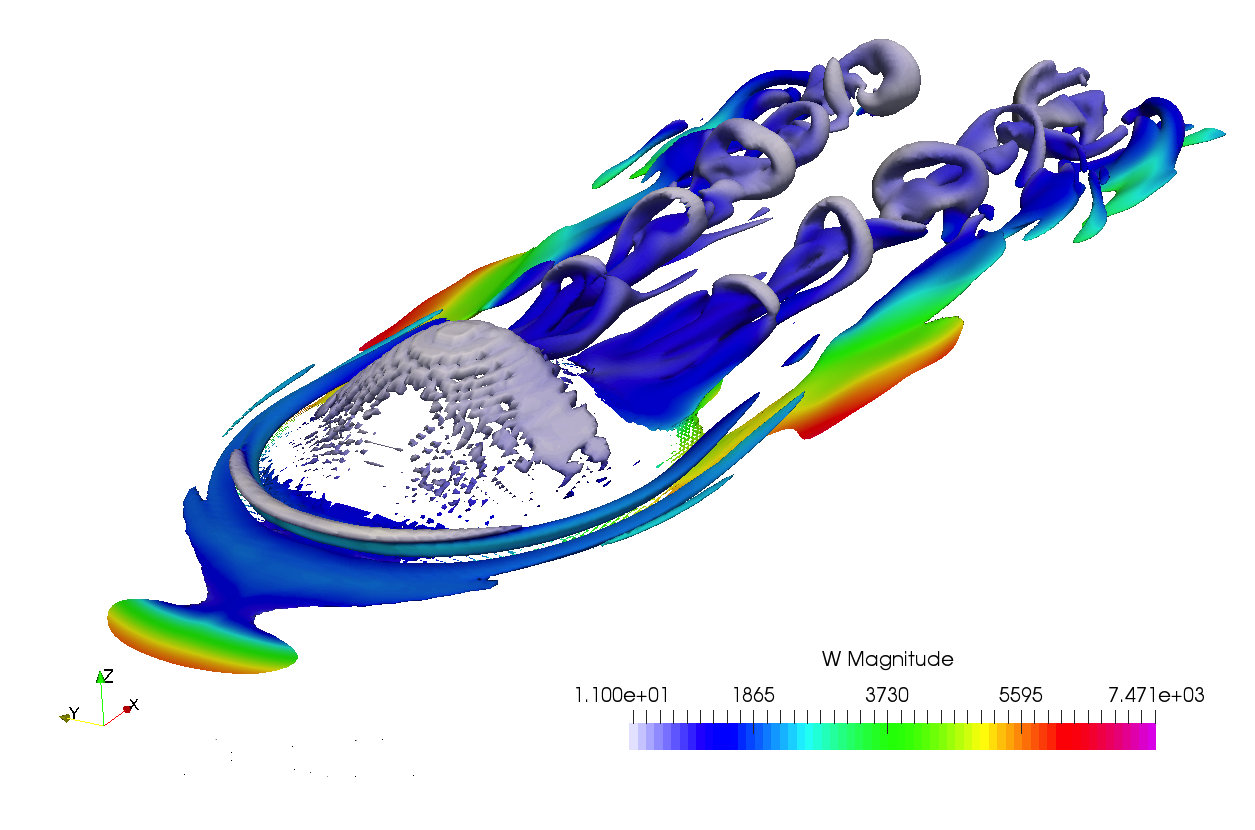}}
\end{subfigmatrix}
 \caption{Topology of hairpin vortices as visualized
by the Q-criterion at $5 \times 10^5$, colored by the values of the gyration. 
Low gyration values for hairpin vortex indicate less variation in
small-scale rotation.}
\label{f:secondInvariantStructures}
\end{figure}
Figure \ref{f:SeparationBubble}
highlights the
formation of the separation bubble on the windward side of the
hill using streamlines of the flow along the centerline $y = 0$. The
leeward side does not show any apparent separtion bubble.
Meanwhile the outline of the MCT windward side bubble falls slightly below some 
of
the recirculation in the MCT data. {\color{black} The reattachment point for the 
MCT data, at
$x = -1.66H$, is slightly downstream than the prediction by Castagna et. al. at $x = -1.7H$, but 
the size of the windward bubble in MCT clearly exceeds the 
predictions by Castagna et. al. Inflow turbulence generated near the wall from the gyration and 
velocity fluctuations may have lead to a much earlier separation point for the 
MCT flow. }

Figure \ref{f:SimpsonComparison} shows that
experimental data for the pressure coefficient obtained from Simpson
\cite{simpson2002study} line up more closely with MCT than the numerical
results from Castagna et. al. \cite{castagna2014direct}\hspace{.1mm}. These successful
comparisons add confidence to the
statement that numerical simulations of MCT produce realistic, physical flow
data without the need for excessively dense grids or high computational costs.
{\color{black} The pressure peak downstream of the hill appears at $x = 2.311H$. 
The DNS study failed to capture this peak, but MCT clearly demonsrates 
a local peak in this region. After this peak, MCT captures the further 
evolution of the wall pressure while the data by Castagna et. al. overpredicts the surface 
pressure. 
The 
rapid differences in the vorticity found near the secondary local peak are 
likely the cause of the local variance in pressure. Simpson noted that the 
pressure coefficient could be directly related to the vorticity flux 
\cite{simpson2002study}, and the resolution of this vorticity near the wall 
likely helps the resolution of the surface pressure. }

Near the wall, turbulent fluctuations in MCT data behave in recognizable
patterns, but contain noticeable differences from Navier-Stokes
simulations. Figure \ref{f:Gyr} compares the turbulence intensities from
Castagna et. al. \cite{castagna2014direct} and MCT
data at the same Reynolds number. {\color{black} The MCT turbulence
intensities are calculated from gyration. The gyration variable
is considered as a stochastic variable and used to perturb the velocity field.
The MCT turbulence intensities ($I$) are obtained by finding the root-mean-square value of gyration \cite{CWC2018}}
\begin{equation}
\omega_\text{rms}=\sqrt{\frac{1}{N}\sum_{k=1}^{N}\omega_i'\omega_i'} \qquad \text{and}\qquad I=\frac{\omega_\text{rms}\sqrt{j}}{u_\tau}
\end{equation}

 {\color{black} Qualtiatively similar results 
are found near the edge of the boundary layer downstream of the hill, and 
near the wall upstream of the hill. MCT results demonstrate the characteristic 
peak and decline in turbulence intensity, with some noticeable differences in 
the
wall-normal data. Downstream of the bump at $x = 11.6H$, the boundary layer in 
the MCT data shrinks considerably, leading to a wider spread of the turbulence 
intensity. The downstream behavior is likely affected by the formation of
structures such as hairpin vortices. }

More in-depth information of the structure of the flow is found using the
MCT Q-criterion \cite{Chen2018PRE}. Figure
\ref{f:secondInvariantStructures} shows the isosurface of the Q-criterion 
colored by the gyration. {\color{black} Figure \ref{f:secondInvariantStructures} 
reveals
structures within the turbulence, and particularly hairpin vortices downstream
of the hill. Periodic regular hairpin vortices emerge for the 
Q-value of $5 \times 10^5$, with the arches of these vortices characterized by 
a low value of gyration. Since these structures are all described by 
the same value for the Q-criterion, these lower regions of gyration 
must correspond to higher values of velocity gradients. Here, the 
macroscale component of the flow is dominant. Near the floor, however, 
gyration plays a more critical role in determining local evolution of 
strucutre near the wall. Overall, the new Q-criterion provides the tool
to visualize the evolution of large-scale structures that form within the flow
and see how small and large scale rotations affect the dynamics and
topology of the bulk flow.}

\section{Conclusion} \label{Conclusion}

Direct comparisons with the experimental results of Simpson
\cite{simpson2002study} and the numerical data of Castagna et. al.
\cite{castagna2014direct} suggest that MCT can reproduce
several aspects of flows simulated using the N-S equations on coarser grids and
provide new tools for visualizing the behavior of flow structures. 
{\color{black} The obtained
plot of the separation bubble lines on the windward side of
the bump yields recirculation qualitatively similar to DNS results, and the 
associated pressure coefficient matches more closely with experimental data 
likely due to the presence of large vorticity differences near the wall. 
Turbulence
intensities are shown to be qualitatively similar to the data by Castagna et. al., though 
differences are found primarily near the wall. The sharp
drop in mesh size as compared with Castagna et. al. \cite{castagna2014direct} shows
turbulent fluctuations, hairpin vortices, and pressure trends can be obtained
without incurring high computational costs.} Furthermore, MCT adds an objective
Q-criterion for flow visualization, and can observe variation in local rotation
in these structures through the gyration variable. Since the gyration acts on
the flow only at the smallest scales, an analysis can be done to see how the
inner structure of the flow affects the behavior of larger structures. The
contour plots suggest that hairpin vortices contain regions of high gyration 
near the wall but develop regions of low gyration at the top. These regions can 
show where small or large scale differences in rotation predominate.

These initial comparisons with DNS and experimental data represent a first step
for MCT to model compressible turbulence and identify vortices. The new
Q-criterion makes possible a new discussion of what variables and which length
scales matter for the formation and evolution of structures within
turbulent flows. Future work with canonical cases
involving the formation of vortices will illuminate how sub-scale motion
affects the formation or decay of large and small vortices. Altering
values for $\kappa$, which denotes the contribution of the inner structure to
the evolution of turbulence will show the role gyration plays in the shape
and stability of vortices. Identifying vortices and distinguishing them from
large or small-scale eddies in the flow will give a better assessment of the
structure of the turbulence, which will in turn yield better information on the
turbulence intensity, pressure, and stress profiles. With the multiscale
character of MCT, this information on the structure of turbulence can be
obtained from smaller mesh sizes and fewer computational resources. {\color{black} It should be 
emphasized that MCT is constructed on the basis of two different length scales within a continuum. On the contrary, RANS and LES are formulated within the classical continuum with single length scale and the additional length scale is introduced through statistical averaging or filtering. Therefore, unlike RANS and LES, the MCT constitutive equations can be completely formulated through the second law of thermodynamics and sufficient to close the system of the governing equations. 
}


\section*{Acknowledgements}

This material is based
upon work supported by the Air Force Office of Scientific Research under award
number FA9550-17-1-0154. The authors would like to thank our coworkers Mohamed Mohsen Ahmed and Allison Adams for their
assistance and insights at every stage of this process. 
\bibliography{bibtex_database}

\end{document}